\documentclass[aps,pre,onecolumn,noshowpacs,superscriptaddress,groupedaddress,preprintnumbers,nofootinbib,showpacs,12pt]{revtex4}

\usepackage{graphicx,amsmath,amssymb,dcolumn,physics,color,braket,bm,cases}

\begin{document}
\title{Stochastic gradient descent introduces an effective landscape-dependent regularization favoring flat solutions}

\author{Ning Yang}
\affiliation{Peking-Tsinghua Center for Life Science, Peking University, Beijing 100871, China}

\author{Chao Tang}
\affiliation{Peking-Tsinghua Center for Life Science, Peking University, Beijing 100871, China}
\affiliation{Center for Quantitative Biology, Peking University, Beijing 100871, China}
\affiliation{School of Physics, Peking University, Beijing 100871, China}

\author{Yuhai Tu}
\affiliation{IBM T. J. Watson Research Center, Yorktown Heights, New York 10598, USA}

%\date{\today}

\begin{abstract}
     Generalization is one of the most important problems in deep learning (DL). In the overparameterized regime in neural networks, there exist many low-loss solutions that fit the training data equally well. The key question is which solution is more generalizable. Empirical studies showed a strong correlation between flatness of the loss landscape at a solution and its generalizability, and stochastic gradient descent (SGD) is crucial in finding the flat solutions. To understand how SGD drives the learning system to flat solutions, we construct a simple model whose loss landscape has a continuous set of degenerate (or near degenerate) minima. By solving the Fokker-Planck equation of the underlying stochastic learning dynamics, we show that due to its strong anisotropy the SGD noise introduces an additional effective loss term that decreases with flatness and has an overall strength that increases with the learning rate and batch-to-batch variation. We find that the additional landscape-dependent SGD-loss breaks the degeneracy and serves as an effective regularization for finding flat solutions.
Furthermore, a stronger SGD noise shortens the convergence time to the flat solutions. However, we identify an upper bound for the SGD noise beyond which the system fails to converge. Our results not only elucidate the role of SGD for generalization they may also have important implications for hyperparameter selection for learning efficiently without divergence.   
\end{abstract}

\maketitle

\clearpage

Deep learning (DL) \cite{lecun_deep_2015} has achieved tremendous success across various fields ranging from image recognition~\cite{he2016deep} to playing Go~\cite{alpha-go}, and even solving complex scientific problems such as protein folding~\cite{AlphaFold}. The parameters (weights) in a neural network model are trained by following gradient decent of a global loss function. Given the large number of parameters in DL, there are many solutions that have the same (or nearly the same) minimum loss. Of course, the ``goodness" of a solution is measured by its generalizability, i.e., its performance in fitting previously unseen testing data, which differs from solution to solution. Indeed, generalization remains the most important problem in DL~\cite{yang_rethinking_2020}. Previous empirical studies showed that generalization correlates with the local curvature of the loss landscape around the solution: flatter solutions tend to generalize better than sharper ones \cite{jastrzebski_three_2018,wu_how_2018,zhang_energyentropy_2018,keskar_large-batch_2017,smith_bayesian_nodate,hochreiter_flat_1997}. Recent work based on an activity-weight duality showed the dependence of the generalization loss on the flatness of loss landscape explicitly~\cite{feng2022activity}. In this paper, we aim to understand why and how the SGD learning algorithm finds flat solutions. 

From previous studies~\cite{sagun_empirical_2018,ghorbani_investigation_2019,wu_how_2018,chaudhari_entropy-sgd_2019}, the loss landscapes are highly degenerate in the overparameterized regime, where the number of network parameters greatly exceeds the number of training samples. This is evidenced from Hessian spectrum analysis: most eigenvalues of the Hessian matrix ($\mathbf{H}$) at the solution are nearly zero; only a few eigenvalues are significantly larger than zero, which means that most directions are relatively degenerate (flat) except for very few non-degenerate (sharp) directions. Furthermore, the structure of the Hessian spectrum is stabilized after only a few epochs\footnote{An epoch is the training time when every sample in the training set has been used once.} of training~\cite{ghorbani_investigation_2019,feng2021phases} and the subspace composed of these degenerate directions are connected without barriers~\cite{garipov_loss_2018,draxler_essentially_2018}. These studies indicate that during most of the training processes, DL searches in a highly degenerate loss landscape instead of escaping from local minima in a rugged landscape~\cite{wei_how_2019}. 

A natural question thus arises: in a degenerate loss landscape with many solutions having almost the same low loss, how does DL pick a flatter one? The answer has to do with stochastic gradient descent (SGD), which was originally adopted due to computational limitation in calculating gradients of the overall loss function with all the training data \cite{robbins_stochastic_1951,bottou_large-scale_2010}. Different from gradient descent (GD) training with all samples, SGD only uses a random minibatch of samples to compute the gradients and update parameters $\bm{\theta}$ at each iteration. The updating rule of SGD for iteration $t$ is given by:
\begin{equation} \label{eq:updating_rule_SGD}
    \bm{\theta}_{t+1}=\bm{\theta}_{t}-\eta \nabla_{\bm{\theta}} L^{\mu}\left(\bm{\theta}_{t}\right), 
\end{equation}
where $\eta$ is the learning rate, $L^{\mu}(\bm{\theta})$ denotes the minibatch loss for a random minibatch $\mu$. Theoretically, the overall loss $L(\bm{\theta})$ for all samples equals to the ensemble average of minibatch loss, i.e., $L(\bm{\theta}) = \Braket{L^{\mu}(\bm{\theta})}_{\mu}$ \cite{feng_inverse_2021}. SGD noise originates from the difference between $L(\bm{\theta})$ and $L^{\mu}(\bm{\theta})$, which is characterized by the covariance matrix $\mathbf{C}(\bm{\theta})$ with its elements given as:
\begin{equation} \label{eq:covariance_definition}
    C_{ij}(\bm{\theta}) = \Braket{ \qty\Big( \pdv{L}{\theta_i}-\pdv{L^{\mu}}{\theta_i} ) \qty\Big( \pdv{L}{\theta_j}-\pdv{L^{\mu}}{\theta_j} )}_{\mu}.
\end{equation}
Previous studies found that SGD noise covariance matrix $\mathbf{C}(\bm{\theta})$ is highly correlated (aligned) with the Hessian matrix $\mathbf{H}(\bm{\theta})$ \cite{zhu_anisotropic_2019,jastrzebski_three_2018,xie_diffusion_2021}, which means that SGD noise is highly anisotropic just as the Hessian itself. 

Besides its anisotropic structure, the SGD noise has an overall strength that depends on the learning rate and the minibatch size~\cite{jastrzebski_three_2018,smith_bayesian_nodate}. In this paper, by using minimal models of SGD and the degenerate loss landscape we aim to understand two general questions in DL: (1) how the anisotropic SGD noise drives the system to flat minima in highly degenerate loss landscapes; (2) how the SGD noise strength affects the final solutions. 

\begin{figure}[htbp] 
    \includegraphics[width=0.8\textwidth]{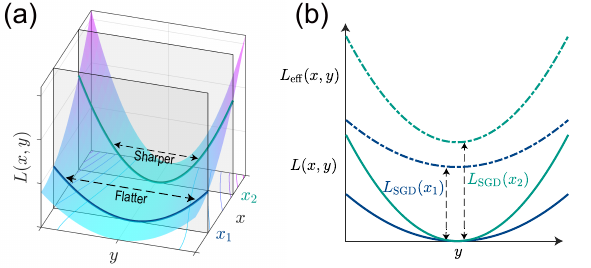}
    \caption{(a) Illustration of two-dimensional degenerate loss function. Each point on $y=0$ is a solution (minimum) with different flatness, the solution at $x_1$ (blue line) is flatter than that at $x_2$ (green line). (b) Loss landscape before (solid line) and after (dash-dotted line) reshaping by anisotropic noise, the landscape of flatter minimum $x_1$ is lower than sharper minimum $x_2$ after reshaping.}
    \label{fig:introduction}
\end{figure}

\emph{Anisotropic noise breaks degeneracy of solutions.} 
To demonstrate effects of the anisotropic SGD noise, we consider the simplest model with a two-dimensional loss function $L(x,y)$ where $x$ and $y$ represent the degenerate (flat) and non-degenerate (sharp) directions, respectively:
\begin{equation} \label{eq:loss_fully_degenerate}
    L(x, y) =   \frac{1}{2} \lambda(x)y^2 = \frac{y^2}{2F^2(x)},
\end{equation}
which has a line of degenerate minima (solutions) at $y=0$ as illustrated in Fig.~\ref{fig:introduction}a. At a given solution $(x,0)$, the flatness along the non-degenerate $y$-direction is defined as the inverse square root of the curvature in the $y$-direction, i.e., $F(x) \equiv 1/\sqrt{\lambda(x)}$, so that a larger $F(x)$ corresponds to a flatter minimum. 

The stochastic dynamics of this system is driven by gradient of loss function and the anisotropic noise, which can be described by the following Langevin equations:
\begin{equation} \label{eq:Langevin}
    \dot{x} = - \partial_x L(x,y) + \xi_1(t); \quad \dot{y} = - \partial_y L(x,y) + \xi_2(t),
\end{equation}
where $\xi_1(t)$ and $\xi_2(t)$ are the noise for $x$ and $y$, respectively. The corresponding dynamics of the probability density $P(x,y,t)$ is governed by the Fokker-Planck equation:
\begin{equation}  \label{eq:Fokker-Planck}
    \begin{split}
        \pdv{P}{t} =  \pdv{x} \left( \pdv{L}{x}  + \pdv{x}D_{11} + \pdv{y}D_{12}  \right)P + \pdv{y} \left( \pdv{L}{y} + \pdv{x} D_{21}  +\pdv{y} D_{22}   \right) P ,        
    \end{split}
\end{equation}
where $\mathbf{D}$ is the covariance matrix of the noise. For simplicity, we first consider the case where $D_{11}=\kappa^{-1}\Delta$ and $D_{22}=\Delta$ with $\Delta$ denoting a constant noise strength in the $y$-direction and $\kappa>1$ characterizing the noise anisotropy; we also assume that $\xi_1$ and $\xi_2$ are uncorrelated, i.e., $D_{12}=D_{21}=0$.  By taking advantage of the separation of timescales in the degenerate and non-degenerate directions, we can    integrate the  fast variable $y$ and solve for the steady-state distribution (see Sec.~\uppercase\expandafter{\romannumeral3}B in SM for the detailed derivation):
\begin{equation} 
    \begin{split}
    P_{\mathrm{ss}}(x, y) = \frac{1}{Z_{\mathrm{ss}}} \exp \left[-\frac{\lambda(x) y^{2}}{2 \Delta}\right] \lambda(x)^{-\frac{\kappa-1}{2}} \equiv \frac{1}{Z_{\mathrm{ss}}}\exp \left[ - \frac{L_\mathrm{eff}(x,y)}{\Delta} \right],
    \end{split}
\end{equation}
where $Z_{\mathrm{ss}}$ is the normalization factor. The effective loss function $L_\mathrm{eff}(x,y)$ is given by:
\begin{equation} \label{eq:effective_potential}
    \begin{split}
        L_\mathrm{eff}(x,y) \equiv -\Delta \ln [Z_{\mathrm{ss}}P_{\mathrm{ss}}(x,y)] 
         = \underbrace{\frac{1}{2} \lambda(x)y^2}_{L(x,y)} + \underbrace{\vphantom{\frac{1}{2}} \Delta (1 - \kappa) \ln{F(x)}}_{ L_\mathrm{SGD}(x)}, \\
    \end{split}
\end{equation}
which shows explicitly that the anisotropic noise reshapes the original loss landscape by introducing an additional flatness-dependent loss $L_\mathrm{SGD}(x) \equiv  \Delta (1-\kappa) \ln{F(x)}$. This additional loss breaks the degeneracy in the original loss function. Since the SGD noise is stronger in the non-degenerate (sharp) direction, i.e., $\kappa >1$, $L_\mathrm{SGD}$ is lower when $F(x)$ is larger and therefore an anisotropic noise with $\kappa >1$ favors the flatter solutions. Encouraged by these results from the simple model, we next consider a more realistic model with a learning dynamics that mimics the realistic SGD and a loss landscape that is not perfectly degenerate. 

\emph{The random-shift minibatch model.} 
In the simple model, the anisotropic noise is put in by hand in the Langevin dynamics (Eq.~\ref{eq:Langevin}). However, SGD noise originates from variations in the minibatch loss function during training. To mimic realistic SGD without considering details of network structure and distribution of train data, we construct an ensemble of minibatch losses by randomly shifting $L(x,y)$.  Specifically, we approximate minibatch loss $L^{\mu}(x,y)$ by taking small random shifts $\mu_1$ and $\mu_2$ in the overall loss $L(x,y)$ and keeping only the first-order expansion term:
\begin{equation} \label{eq:loss_minibatch_2D}
    \begin{split}
        L^{\mu}(x,y) = L(x-\mu_1, y-\mu_2)
        \approx L(x,y) - \pdv{L(x,y)}{x} \mu_1 - \pdv{L(x,y)}{y} \mu_2,   
    \end{split}
\end{equation}
where we assume the shifts are independent white noise with zero mean and noise strength $2\sigma$. Empirically, noise strength $\sigma$ depends inversely on the batch size, and $\sigma=0$ corresponds to the case where batch size is equal to whole training data size, which corresponds to GD. Under this construction, we can show that the noise covariance matrix $\mathbf{C}$ is directly related to the Hessian matrix: $\mathbf{C} = 2\sigma \mathbf{H}^2$ (see Sec.~\uppercase\expandafter{\romannumeral1}A in SM). It is worth emphasizing that there are other ways to construct the minibatch loss ensemble, which could lead to a different form of the $\mathbf{C}$-$\mathbf{H}$ dependence. However, as long as they are positively correlated (aligned), which leads to anisotropic noise that is stronger in the sharper direction as observed in realistic SGD~\cite{feng_inverse_2021}, our general conclusions remain valid independent of the exact form of their dependence.  

Another assumption made in the simple model is that the overall loss landscape is perfectly degenerate. Here, we consider a more realistic loss landscape where the flatness in the $x$-direction is large but not infinite. Specifically, we modify Eq.~\ref{eq:loss_fully_degenerate} by adding a term $\frac{1}{2} \varepsilon \lambda_0 x^2$ with $0<\varepsilon \ll 1$: 
\begin{equation} \label{eq:loss_highly_degenerate}
    L(x, y) =  \frac{1}{2} \varepsilon \lambda_0 x^2 + \frac{1}{2} \lambda(x)y^2,
\end{equation}
where we expand $\lambda(x)=\sum_{n=0}\lambda_{n}x^{2n}$ around the global minimum $(x^{*}=0,y^{*}=0)$ with $\lambda_n$ the $2n$th-order expansion coefficient. Given $\varepsilon\ll 1$, points along the valley ($y=0$) near $x=0$ can be considered as near degenerate solutions with low loss. Among these solutions, there are two special ones: a global minimum $x^{*}$ defined by the minimum of $L$ and the flattest solution $x_{F}$ given by the maximum of flatness function $F(x)$. If the two locations coincide: $x^*=x_{F}=0$, SGD will find the global minimum (also the flattest solution) and increasing the SGD noise has the advantage of reaching the solution faster (see Fig.~S2 in SM). However, if the flattest solution is not the same as the global minimum, e.g., when $\lambda_1 < 0, \lambda_2 > 0$, $\lambda_n =0 $ for $n>2$, which solution does the SGD find? And how does it depend on the SGD noise strength? We answer these two questions next. 
\begin{figure}[htbp] 
    \includegraphics[width=0.8\textwidth]{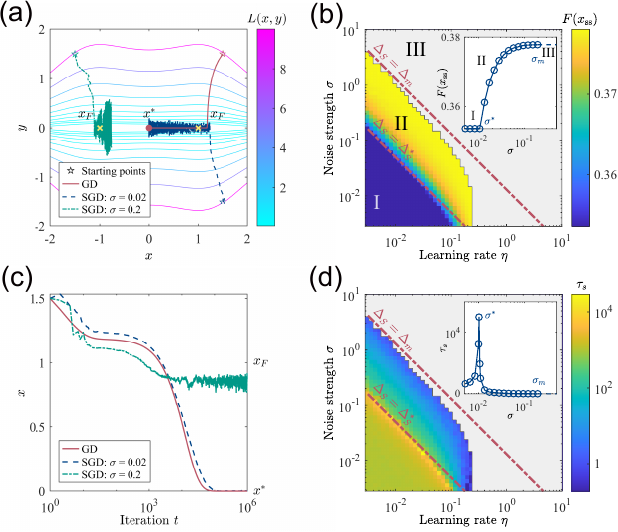}
    \caption{(a) Trajectories of SGD on a highly degenerate loss with fixed learning rate $\eta=0.01$ and different noise strength: $\sigma=0$ (red solid line), $\sigma=0.02$ (blue dashed line), $\sigma=0.2$ (green dash-dot line). Starting point is $(1.5,1.5)$, marked on symmetrical positions for visualization. Global minimum $x^{*}=0$ and flattest solution $x_{F}=\pm1$ are marked as red point and yellow cross, respectively. The loss landscape parameters used in simulation: $\varepsilon = 0.001$, $\lambda_0 = 8$, $\lambda_1 = -2$, $\lambda_2 = 1$. (b) ($\sigma,\eta$) phase diagram of the flatness of steady-state solution $F(x_\mathrm{ss})$.  (c) Projections of the trajectories onto the degenerate direction $x$ versus iteration $t$. (d) ($\sigma,\eta$) phase diagram for search time $\tau_s$. In (b)\&(d), the two transition boundaries between the three phases (I, II, III) are marked by red dotted lines; the insets show $F(x_\mathrm{ss})$ and $\tau_s$ versus $\sigma$ for a fixed $\eta=0.04$ with the critical $\sigma$ and maximum $\sigma$ denoted by $\sigma^{*}$ and $\sigma_m$, respectively.}
    \label{fig:simulation} 
\end{figure}

\emph{The effects of SGD noise.} By using the minibatch loss function (Eq.~\ref{eq:loss_minibatch_2D}) with the overall loss function (Eq.~\ref{eq:loss_highly_degenerate}), we simulate the learning dynamics of $x$ and $y$ according to SGD updating rules (Eq.~\ref{eq:updating_rule_SGD}). For a fixed learning rate $\eta$, the final steady-state solution $x_\mathrm{ss}$ depends on the noise strength $\sigma$, as shown in Fig.~\ref{fig:simulation}a. In the parameter space spanned by $(\sigma,\eta)$, there are three distinct phases characterized by the flatness of the solution, as shown in Fig.~\ref{fig:simulation}b: for small $\sigma$ and $\eta$, the steady-state solution is always the (sharper) global minimum $x^{*}$, i.e., $F(x_\mathrm{ss})=F(x^{*})$ (Phase \uppercase\expandafter{\romannumeral1}); for intermediate $\sigma$ and $\eta$, flatter solutions are found with $F(x_\mathrm{ss})$ an increasing function of $\sigma$ and $\eta$ (Phase \uppercase\expandafter{\romannumeral2}); for very large $\sigma$ and/or $\eta$, the system diverges (Phase \uppercase\expandafter{\romannumeral3}). The inset of Fig.~\ref{fig:simulation}b shows the three phases as we vary $\sigma$ for a fixed $\eta=0.04$ where $\sigma^{*}$ and $\sigma_m$ represent the transition boundary between the three phases. 

The rate of convergence during training also depends on the SGD noise strength as shown in Fig.~\ref{fig:simulation}c. To quantify the convergence rate, we define the search time $\tau_s \equiv \eta\Braket{t_s}$ as the average first passage time $t_s$ of reaching the solution $x=x_{\mathrm{ss}}$ multiplied by the learning rate $\eta$ ($\eta$ serves as time interval). Analogous to critical slowing down \cite{strogatz_nonlinear_2018}, the search time is much longer near the transition point $\sigma^{*}$ as shown in the inset of Fig.~\ref{fig:simulation}d. Beyond the transition point ($\sigma>\sigma^*$), increasing the noise strength $\sigma$ drastically speeds up the training processes until $\sigma$ approaches its upper limit $\sigma_m$ when the system diverges. 

Changing the learning rate $\eta$ leads to a similar transition in $F(x_\mathrm{ss})$ and $\tau_s$ as changing the noise strength $\sigma$, as shown in Fig.~\ref{fig:simulation}b\&d. In fact, the behavior of the system in the convergence regime (Phase \uppercase\expandafter{\romannumeral1}\&\uppercase\expandafter{\romannumeral2}) is roughly determined by the product of these two hyperparameters $\eta\sigma$, therefore, we define $\Delta_S \equiv \eta\sigma$ as the effective SGD noise level. The transition boundary $\Delta_S=\Delta_S^{*}$ can be approximated by $\Delta_S^*=-\varepsilon/\lambda_1$ (marked with the lower red line). For very small $\Delta_S$ below the transition boundary, the final solution always converges to the sharper global minimum with a moderate convergence rate; when increasing $\Delta_S$ until close to the transition boundary, the convergence rate becomes much slower; and finally for large $\Delta_S$ beyond the transition boundary, a higher $\Delta_S$ would drive the system to flatter solutions with much faster speed. 

However, excessive SGD noise causes the system to diverge~\cite{wu_how_2018}. Explicitly, the convergence condition of the system satisfies: $\eta\Braket{\lambda(x)}_\mathrm{ss}<2$ with $\Braket{\lambda(x)}_\mathrm{ss}\approx \lambda(x_\mathrm{ss})+\frac{1}{2}\lambda''(x_\mathrm{ss})\left( \Braket{x^2}_\mathrm{ss} - x^2_\mathrm{ss}\right)$, which increases with $\Delta_S$ ($\Braket{\cdot}_\mathrm{ss}$ denotes average over the steady-state distribution, see Sec.~\uppercase\expandafter{\romannumeral2}B in SM for details). Therefore, there is a maximum noise level $\Delta_m$, beyond which the system fails to converge. The additional factor $\eta$ in the convergence condition makes the dependence on $\sigma$ and $\eta$ asymmetric, i.e., for the same level of SGD noise ($\Delta_S$), larger learning rate $\eta$ makes the system more likely to diverge, which is consistent with numerical results shown in Fig.~\ref{fig:simulation}b\&d. 

%Overall, our results suggest that the learning rate and the batch size should be chosen to create a large SGD noise level within the convergence regime ($\Delta_S < \Delta_m$) in order to find flat solutions efficiently. 

\emph{An effective flatness-dependent regularization due to SGD noise.} 
To understand the numerical results presented above, we next follow previous work \cite{jastrzebski_three_2018,zhu_anisotropic_2019,chaudhari_stochastic_2018,li_stochastic_nodate} to study the continuous-time stochastic differential equation for the discrete-time SGD updates in the limit of small learning rate, which has the same form as the Langevin equations given in Eq.~\ref{eq:Langevin}. After continuous-time approximation, the noise covariance matrix $\mathbf{D}$ can be determined exactly for the random-shift minibatch loss function ensemble: $\mathbf{D}= \Delta_S\mathbf{H}^2$ (see Sec.~\uppercase\expandafter{\romannumeral3}A in SM for details). Plugging the explicit expression of $\mathbf{D}$ in the Fokker-Plank equation (Eq.~\ref{eq:Fokker-Planck}), the steady-state distribution $P_\mathrm{ss}(x,y)$ can be solved approximately by integrating out the fast variable $y$. To characterize the SGD noise, we define an effective noise strength $\Delta(x)\equiv D_{22} \approx \Delta_S \lambda(x)^2$ and an effective anisotropy $\kappa(x)\equiv \Delta(x) / \Braket{D_{11}}_y$ where we use $\Braket{\cdot}_y$ to denote integration over $y$ for a fixed $x$. Substituting the definition of effective loss function, i.e., $L_\mathrm{eff}(x,y) \equiv -\Delta(x) \ln P_{\mathrm{ss}}(x,y)$, we finally obtain the expression for the additional SGD loss $L_\mathrm{SGD}(x)\equiv  L_\mathrm{eff}(x,y) - L(x,y)$ in terms of $\Delta(x)$ and $\kappa(x)$ (see Sec.~\uppercase\expandafter{\romannumeral3}B in SM for details):
\begin{equation} 
    L_\mathrm{SGD}(x) = \Delta(x) \int{ \left[ \frac{\kappa(x)\varepsilon\lambda_0 x}{\Delta(x)} + \frac{\kappa(x)\lambda'(x)}{2\lambda(x)} \right] \dd{x}} -\frac{\Delta(x)}{2} \ln{\frac{\lambda(x)}{\Delta(x)} - \frac{1}{2}\varepsilon \lambda_0 x^2} .     
    \label{eq:loss_SGD}   
\end{equation}
Form Eq. \ref{eq:loss_SGD}, the critical noise level can be determined: $\Delta_S^{*}\equiv-\varepsilon/\lambda_1$, which agrees with the simulation results. The maximum noise level to avoid divergence can also be obtained approximately: $\Delta_m\approx \sqrt{2/(3\eta_\mathrm{min} \lambda_1^2)}F(x_F)$ with $\eta_\mathrm{min}$ the minimum learning rate, which shows that flatter landscape can tolerate larger SGD noise (see Sec.~\uppercase\expandafter{\romannumeral3}B in SM for detailed derivation of $\Delta_m$). 

In the vicinity of global minimum where the noise level $\Delta_S$ is small enough to ensure the average loss $\Braket{L}\ll 1$,  $\kappa(x)$ and $\Delta(x)$ can be approximated as constants, i.e., $\Delta \approx \Delta_S\lambda_0^2 \ll 1$ and $\kappa \approx (\varepsilon^2 + 2\varepsilon\Delta_S\lambda_1 + 3\Delta_S^2\lambda_1^2)^{-1}\gg 1$, and $L_\mathrm{SGD}(x)$ can be expressed as:
\begin{equation} \label{L_SGD}
    \begin{split}
    L_\mathrm{SGD}(x) \approx  \Delta (1 - \kappa) \ln{F(x)} +(\kappa-1)L(x,y=0),
    \end{split}
\end{equation}
which contains the flatness-dependent term $L_\mathrm{SGD,F}\equiv\Delta (1 - \kappa) \ln{F(x)}$ that decreases with $F(x)$ and another loss-dependent term $L_\mathrm{SGD,L}\equiv(\kappa-1)L(x,y=0)$, which is proportional to the original loss function $L(x,y=0)$ at the nearly degenerate solutions\footnote{The second term is absent in the purely degenerate case when $\varepsilon=0$, see Eq.~\ref{eq:effective_potential}.}. %$Note that integrating out the fast variable in the sharp direction also introduces the second term in Eq.~\ref{L_SGD}, which is proportional to the original loss function $L(x,y=0)$ at the nearly degenerate solutions\footnote{The second term is absent in the purely degenerate case when $\varepsilon=0$.}.   
Eq.~\ref{L_SGD} shows clearly how $L_\mathrm{SGD,F}$ compete with $L_\mathrm{SGD,L}$: after integrating out the fast variable in the sharp direction, both $L_\mathrm{SGD,F}$ and $L_\mathrm{SGD,L}$ are affected by the anisotropy $\kappa$, however, the flatness-dependent SGD loss $L_\mathrm{SGD,F}$ has an overall strength proportional to the SGD noise level $\Delta_S$. Therefore, increasing $\Delta_S$ would increase the contribution of the flatness-dependent SGD loss $L_\mathrm{SGD,F}$, which serves as an effective regularization that favors the flatter solutions.

\begin{figure}[htbp] 
    \includegraphics[width=0.8\textwidth]{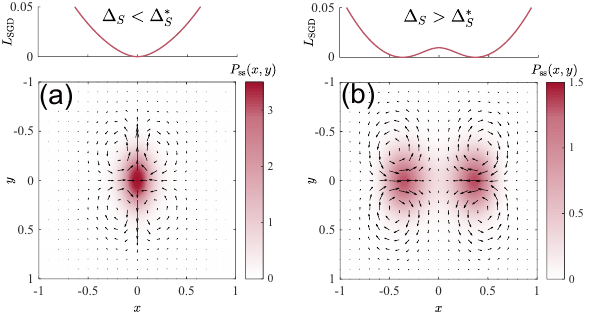}
    \caption{Numerical solutions of steady-state distribution $P_\mathrm{ss}(x,y)$ (contour plot) and the probability fluxes $(J_x, J_y)$ (vector field) for $\Delta_S=0.009$ (a) and $\Delta_S=0.012$ (b), respectively. Loss landscape parameters: $\varepsilon = 0.02$, $\lambda_0 = 8$, $\lambda_1 = -2$, $\lambda_2 = 1$ and transition boundary $\Delta_S^{*}=-\varepsilon/\lambda_1=0.01$. Corresponding $L_\mathrm{SGD}$ are plotted on the top panel.}
    \label{fig:analytical}
\end{figure}

\emph{The nonequilibrium learning dynamics in SGD.} Instead of using the adiabatic approximation to obtain an approximate expression for the projected SGD loss $L_\mathrm{SGD}(x)$, we have also computed the steady-state distribution $P_\mathrm{ss}(x,y)$ in the full $(x,y)$ plane by solving the Fokker-Planck equation numerically (see Sec.\uppercase\expandafter{\romannumeral3}C in SM for details). From $P_\mathrm{ss}(x,y)$, we can determine the probability fluxes, $J_x \equiv  - \pdv{L}{x} P -  \pdv{x} D_{11}P - \pdv{y} D_{12} P$ and $J_y \equiv - \pdv{L}{y} P - \pdv{x} D_{21} P - \pdv{y} D_{22} P$, in the $x$ and $y$ direction, respectively. In an equilibrium system with isotropic noise, there is no persistent flux in the steady state, i.e., $J_x=J_y=0$. However, as shown in Fig.~\ref{fig:analytical}a\&b, there are non-zero probability fluxes in the steady state, which clearly demonstrates the non-equilibrium nature of the SGD learning dynamics due to the anisotropic SGD noise. The probability fluxes form persistent cycles surrounding the final steady-state solution. In the high SGD noise case when $\Delta_S >\Delta_S^*$, these nonequilibrium flux cycles are crucial for stabilizing the flat solutions, which are not at the minimum of the original loss function. 

\emph{Discussions.} Generalization is a fundamental problem in DL. Increasing empirical and theoretical evidence show that a flatter solution has a better generalization performance. In this work, we ask how SGD based algorithm drives the learning system to flatter solution in a highly degenerate loss landscape typical in overparameterized neural network models. Our findings indicate that the anisotropic SGD noise introduces an additional flatness-dependent loss, which serves as an implicit regularization that favors the flatter solutions. The hyperparameters in SGD, learning rate $\eta$ and batch noise level $\sigma$, together determine an overall SGD noise strength. A higher SGD noise strength within the convergence bound ($\Delta_s<\Delta_m$) not only allows the system to find flatter solutions but also speed up the search process. Our work explains how SGD helps find flatter solutions it may shed light on the selection of hyperparameters as well as design of more efficient learning algorithms. 

\bibliography{SGD}
\end{document}

\title{SUPPLEMENTARY MATERIALS} 

\author{Ning Yang}
\affiliation{Peking-Tsinghua Center for Life Science, Peking University, Beijing 100871, China}

\author{Chao Tang}
\affiliation{Peking-Tsinghua Center for Life Science, Peking University, Beijing 100871, China}
\affiliation{Center for Quantitative Biology, Peking University, Beijing 100871, China}
\affiliation{School of Physics, Peking University, Beijing 100871, China}
\author{Yuhai Tu}
\affiliation{IBM T. J. Watson Research Center, Yorktown Heights, New York 10598, USA}

\maketitle

\tableofcontents
\newpage

\section{Supplementary materials of model construction}
\subsection{Derivation of the relation between noise covariance matrix and Hessian matrix}

Due to minibatch sampling, there are differences between the minibatch loss function and the overall loss function, which is the origin of SGD noise. In real training process, the exact form of the minibatch loss function is determined by the network structure as well as distribution of train data, which varies case by case. Nevertheless, in our phenomenological two-dimensional model, we want to capture the main characteristics of SGD noise, i.e., landscape-dependency and anisotropy, thus we ignore all other details and construct the random-shift minibatch loss function in Eq.~8 by taking small random shifts from the overall loss function. Here, we'll show this construction indeed introduces an anisotropic landscape-dependent noise as in the real case.

Without loss of generality, we first extend this construction to the N-dimensional variable space $\bm{\theta} \in \mathbb{R}^{N}$ (the case in the main text corresponds to $N=2$), the random-shift minibatch loss function is given by:
\begin{equation} \label{eq:loss_minibatch}
     L^{\mu}(\bm{\theta}) = L(\bm{\theta}-\bm{\mu}) \approx L(\bm{\theta}) - \bm{\mu} \cdot \nabla L(\bm{\theta}),
\end{equation}
where we assume the shift vector $\bm{\mu}$ is the N-dimensional isotropic Gaussian white noise with zero mean and constant noise strength for each dimension, i.e., $\Braket{\mu_i} = 0$ and $\Braket{\mu_i \mu_j} = 2\sigma \delta_{ij}, \forall i,j\in \{1,\cdots,N\}$. $\sigma$ denotes the noise strength which determines how far the minibatch loss is away from the overall loss, corresponding to the inverse of batch size in real SGD. It's easy to check that the ensemble average of minibatch loss $\langle {L^{\mu}(\bm{\theta})} \rangle_{\mu}$ gives the overall loss $L(\bm{\theta})$, which ensures that zero  noise case degenerates to GD: 
\begin{equation} \label{eq:loss_ensemble_average}
    \langle L^{\mu}(\bm{\theta}) \rangle_{\mu} = \langle L(\bm{\theta}-\bm{\mu}) \rangle_{\mu} \approx L(\bm{\theta}) - \Braket{\bm{\mu}} \cdot \nabla L(\bm{\theta}) = L(\bm{\theta}).
\end{equation}

From Eq.~\ref{eq:loss_minibatch}, we then compute the gradients of minibatch loss for $i$th variable:
\begin{equation}
    \begin{split}
        \pdv{L^{\mu}(\bm{\theta})}{\theta_i} &= \pdv{L(\bm{\theta})}{\theta_i} - \pdv{\theta_i} \left( \sum_{j} \pdv{L(\bm{\theta})}{\theta_j} \mu_j \right) \\
        &= \pdv{L(\bm{\theta})}{\theta_i} - \sum_{j} H_{ij}(\bm{\theta}) \mu_j,      
    \end{split}
\end{equation}
where $H_{ij}(\bm{\theta})$ is the $(i,j)$ entry of Hessian matrix $\mathbf{H(\bm{\theta})}$. According to the definition (Eq.~2 of the main text), the $(i,j)$ element of the noise covariance matrix $\mathbf{C(\bm{\theta})}$ is given by:
\begin{equation} \label{eq:covariance_general}
    \begin{split}
        C_{ij}(\bm{\theta}) &\equiv \Braket{ \qty\Big( \pdv{L(\bm{\theta})}{\theta_i}-\pdv{L^{\mu}(\bm{\theta})}{\theta_i} ) \qty\Big( \pdv{L(\bm{\theta})}{\theta_j}-\pdv{L^{\mu}(\bm{\theta})}{\theta_j} )}_{\mu}\\ 
        &= \sum_{kl} H_{ik}(\bm{\theta})H_{jl}(\bm{\theta})\Braket{\mu_k \mu_l} \\
        & = 2\sigma \sum_k  H_{ik}(\bm{\theta})H_{jk}(\bm{\theta}), \\       
    \end{split}
\end{equation}
thus we get the quadratic $\mathbf{C}$-$\mathbf{H}$ relation $\mathbf{C} = 2\sigma\mathbf{H}^2$. Due to the highly degenerate loss landscape, this quadratic alignment indeed introduces a highly anisotropic noise with its strength controlled by the hyperparameter $\sigma$. Note that here we directly assume that the SGD noise is Gaussian white noise, however, in real case, the Gaussian assumption holds when the batch size is large enough due to the central limit theorem \cite{zhu_anisotropic_2019,mandt_stochastic_nodate} and the white assumption holds if we sample minibatches with replacement.

\subsection{Construction of the general form of the overall loss function}

In Eq.~3 of the main text, we directly introduce the simplest case of two-dimensional overall loss function with a perfectly (completely) degenerate valley. In general, for high-dimensional case where there exists a region of connected degenerate minima, suppose there are $n$ non-degenerate variables and $N-n$ degenerate variables, it has been proved that (Lemma 1 in \cite{wei_how_2019}), this kind of loss landscape with a minimal valley can be expanded along the non-degenerate variables under certain transformation and translation: 
\begin{equation} \label{eq:loss_general}
    L(\bm{\theta})=L^{*}+ \frac{1}{2} \sum_{i=1}^{n}  \lambda_{i}\left(\theta_{n+1}, \ldots, \theta_{N}\right) \theta_{i}^{2},
\end{equation}
where $L^{*}$ denotes a minimal constant defined by the minimal valley and $(\theta_1, \dots, \theta_n )$ and $(\theta_{n+1}, \dots, \theta_{N})$ denote non-degenerate and degenerate subspaces respectively, $\lambda_i$s are the positive eigenvalues of the Hessian for the non-degenerate space and depend on the position in the degenerate space. Intuitively, there is a timescale separation between non-degenerate and degenerate subspaces due to the degeneracy, meaning the dynamics of non-degenerate variables has relaxed to the stationary state from very early stage of training, thus we can keep only the second order terms of non-degenerate variables. For the rest of training process, what we observe mostly are the slow dynamics in degenerate subspaces therefore the eigenvalues are function of the degenerate variables. The two-dimensional form in Eq.~3 corresponds to a specific case where $L^{*}=0$, $n=1$ and $N=2$ in Eq.~\ref{eq:loss_general}, which means we keep only one degenerate and one non-degenerate variable.

If we further consider the incompletely degenerate form of overall loss function, we should add the contribution of degenerate variables
in Eq.~\ref{eq:loss_general}, as Eq.~9 of the main text for the two-dimensional case. This contribution is extremely weak (we use a dimensionless constant $\varepsilon$ to quantify it) and we can keep only the second-order term. In such incompletely degenerate landscape, all solutions along $y=0$ are almost equally low with only one global minimum $(x^{*}=0,y^{*}=0)$. To further study the flatness of solutions, we first compute the corresponding Hessian matrix of loss function in Eq.~9 of the main text:
\begin{equation} \label{eq:Hessian_matrix}
    \mathbf{H}(x,y)=\begin{bmatrix}
           \pdv[2]{L(x,y)}{x} & \pdv{L(x,y)}{x}{y} \\
           \pdv{L(x,y)}{y}{x} & \pdv[2]{L(x,y)}{y} \\
        \end{bmatrix} = \begin{bmatrix}
            \varepsilon\lambda_0 + \frac{1}{2} \lambda''(x)y^2 & \lambda'(x)y \\
            \lambda'(x)y & \lambda(x) \\
        \end{bmatrix}.
\end{equation}
The mean curvature of solutions along $y=0$ is proportional to the trace of the Hessian, i.e., $\trace(\mathbf{H}(x,0))=\varepsilon\lambda_0 + \lambda(x)\approx\lambda(x)$. The smaller is the curvature, the flatter is the solution, thus we can define a flatness function $F(x)$ as the inverse square root of $\lambda(x)$, i.e, $F(x)\equiv 1/\sqrt{\lambda(x)}$, under this definition, $F(x)$ has the same unit as variable $y$. The flattest solution $x_F$ corresponds to the maximum of $F(x)$, which is also the minimum of $\lambda(x)$. 

\begin{figure}[htbp] 
    \includegraphics[width=0.8\textwidth]{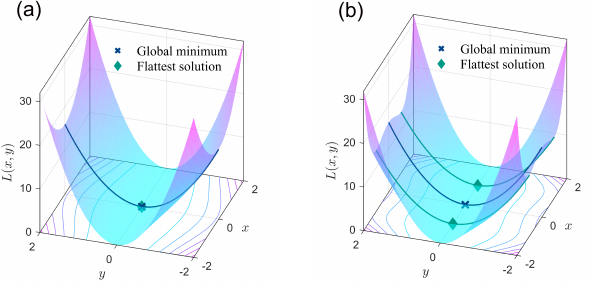}
    \caption{Illustration of loss landscape $ L(x,y)=\frac{1}{2}\varepsilon \lambda_0 x^2 + \frac{1}{2}(\lambda_0 + \lambda_1 x^2 + \lambda_2 x^4)y^2$. (a) The case when global minimum $x^*$ coincides with $x_F$. Global minimum and the flattest solution are marked by blue cross and green diamonds, respectively. The landscape coefficients: $\varepsilon = 0.001$, $\lambda_0 = 8$, $\lambda_1 = 2$, $\lambda_2 = 0$. (b) The case when global minimum $x^*$ doesn't coincide with $x_F$. In this case, the global minimum $x^{*}=0$ and the flattest solutions $x_F=\pm1$. The landscape coefficients: $\varepsilon = 0.001$, $\lambda_0 = 8$, $\lambda_1 = -2$, $\lambda_2 = 1$. }
    \label{fig:loss_landscape}
\end{figure}

Next, we expand $\lambda(x)= \sum_{n=0}^{\infty} \lambda_n x^{2n}$ around the global minimum $x^*=0$ where $\lambda_n$ is $2n$th-order expansion coefficient as in Eq.~9 of the main text. The explicit location of the flattest solution is determined by these coefficients. If the flattest solution coincides with global minimum, i.e., $x_F=x^*=0$, one viable choice of coefficients satisfies $\lambda_1>0$ and $\lambda_n=0$ for $n>1$, the illustrated landscape is shown in Fig.~\ref{fig:loss_landscape}a. For the case where the flattest solution doesn't overlap with the global minimum, we require that $\lambda(x)$ is concave at $x=0$, i.e., $\lambda''(0)=\lambda_1<0$, and to ensure $\lambda(x)$ is always positive, we need to introduce higher-order terms, the simplest form is $\lambda_2>0$, $\lambda_n=0$ for $n>2$, which gives the double-well form we used in the main text, the landscape is shown in Fig.~\ref{fig:loss_landscape}b. In particular, the flattest solutions $x_{F}$ here are located symmetrically around $x=0$ (green diamonds in Fig.~S1), given by $x_{F}=\pm \sqrt{\frac{-\lambda_1}{2\lambda_2}}$. Also note that $\lambda(x_F)>0$ requires $\lambda_0 > \frac{\lambda_1^2}{4\lambda_2}$. 

\section{Supplementary materials of simulation results}
\subsection{The case when global minimum coincides with the flattest solution}
\begin{figure}[htbp] 
    \includegraphics[width=0.8\textwidth]{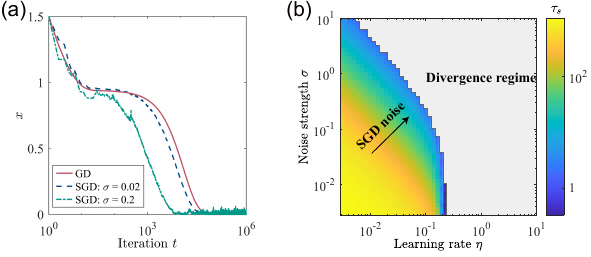}
    \caption{(a) Trajectories of SGD projected onto degenerate direction $x$ with fixed learning rate $\eta=0.01$ and different noise strength $\sigma=0$ (red solid line), $\sigma=0.02$ (blue dashed line), $\sigma=0.2$ (green dash-dot line) v.s. iterations $t$. The landscape coefficients are the same as in Fig.~\ref{fig:loss_landscape}a. (b) Search time $\tau_s$ varying learning rate $\eta$ and noise strength $\sigma$. Increasing SGD noise would speed up the searching process. The divergence regime is marked by shaded area.}
    \label{fig:simulation_quadratic}
\end{figure}

Here we show the simulation results of SGD trajectories of the loss landscape where global minimum coincides with the flattest solution. As depicted in Fig.~\ref{fig:simulation_quadratic}a, when fixing learning rate $\eta$ and varying noise strength $\sigma$, the trajectories converge to the global minimum (also the flattest solution) with different convergence speed. In particular, larger $\sigma$ requires fewer iterations to converge (compare green line with red line in Fig.~\ref{fig:simulation_quadratic}a), thus increasing noise strength $\sigma$ would speed up the convergence. Note that the speed-up effect occurs after a quick descent into the degenerate space (about 10 iterations from the start of training), meaning that SGD can accelerate training process in the highly-degenerate valley of loss landscape. 

As for increasing learning rate $\eta$, the speed-up effect is similar as changing $\sigma$ (Fig.~\ref{fig:simulation_quadratic}b), as we have claimed in the main text that the effective SGD noise is determined by the product of these two hyperparameters $\Delta_S\equiv\eta\sigma$. Here we use the search time $\tau_s\equiv \eta\Braket{t_s}$ rather than average first-passage iterations $\Braket{t_s}$ to quantify the convergence time. The reason is that increasing learning rate $\eta$ would result in fewer iterations thus we should exclude this effect by multiplying $\eta$. Besides, this definition corresponds to absolute time after continuous-time approximation (see Sec.~\ref{sec:continuous_time}). Intuitively, the speed-up effect is due to larger net gradients provided by larger SGD noise, while for GD, the gradients are fully determined by the term $\frac{1}{2}\varepsilon \lambda_0 x^2$, which are relatively small since $\varepsilon \ll 1$.

\subsection{Numerical results of fluctuations and convergence conditions} \label{sec:fluctuation}
In the main text, we have shown that for the case when the global minimum doesn't overlap with the flattest solution, increasing SGD noise results in convergence to flatter solutions and faster convergence speed. Nevertheless, increasing SGD noise would also bring larger fluctuations and higher probabilities of divergence, analogous to the increase in temperature in the equilibrium system. In this section, we'll first show the simulation results of fluctuations and then analyze the convergence conditions.

\begin{figure}[htbp] 
    \includegraphics[width=0.8\textwidth]{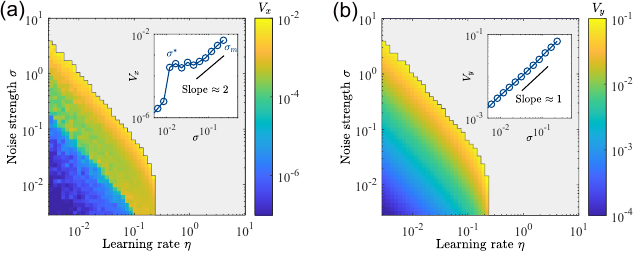}
    \caption{(a) and (b) Steady-state fluctuations $V_x\equiv\Braket{x^2}_\mathrm{ss}-x^2_\mathrm{ss}$ and $V_y\equiv\Braket{y^2}_\mathrm{ss}-y^2_\mathrm{ss}$ varying learning rate $\eta$ and noise strength $\sigma$. The landscape coefficients are the same as in Fig.2 of the main text. The divergence regime is marked with shaded area. Inset: $V_x$ and $V_y$ varying $\sigma$ with fixed $\eta=0.037$.}
    \label{fig:fluctuation}
\end{figure}

We simulate the steady-state fluctuations (variances) $V_x\equiv\Braket{x^2}_\mathrm{ss}-x^2_\mathrm{ss}$ and $V_y\equiv\Braket{y^2}_\mathrm{ss}-y^2_\mathrm{ss}$ numerically varying $\eta$ and $\sigma$. As shown in Fig.~\ref{fig:fluctuation}a\&b, the behaviors of fluctuations for $x$ and $y$ when increasing SGD noise level are different. In particular, while learning rate $\eta$ is fixed, the fluctuation $V_y$ of non-degenerate variable $y$ increases with noise strength $\sigma$ linearly (inset of Fig.~\ref{fig:fluctuation}b). However, the fluctuation $V_x$ of degenerate variable $x$ is more sensitive to SGD noise: the fluctuation increases with noise strength $\sigma$ drastically until the local maximum at critical point $\sigma^{*}$, while for large noise strength (far beyond $\sigma^{*}$), the fluctuation $V_x$ increases with $\sigma$ quadratically (inset of Fig.~\ref{fig:fluctuation}a).

To explain why larger fluctuations make the system more prone to divergence, we shall analyze the convergence (dynamical stability) conditions of SGD. Since the updated gradients are much larger for non-degenerate variable $y$, the stability conditions for variable $y$ should be satisfied to ensure the convergence of the system. Specifically, the SGD updating rule of $y$ for iteration $t$ is given by:
\begin{equation}
    \begin{split}
        y_{t+1} &= y_{t} - \eta \pdv{L^{\mu}(x,y)}{y} \\
        & = y_{t} - \eta \left[  \lambda(x_t)y_t + \lambda'(x_t)y_t\mu_1 + \lambda(x_t) \mu_2 \right] \\
        & = \left[ 1-\eta\lambda(x_t) + \eta\lambda'(x_t) \mu_1 \right]y_t + \eta\lambda(x_t)\mu_2,
    \end{split}
\end{equation}
where we have used the definition of minibatch loss function in the main text. For large enough iteration $t$ where the system has been in the steady-state, the temporal average of $y$ is approximated by:
\begin{equation} \label{eq:first_order_stable}
    \Braket{y_{t+1}}_t \approx \left[ 1 - \eta \Braket{\lambda(x)}_\mathrm{ss} \right] \Braket{y_{t}}_t,
\end{equation}
where $\Braket{\lambda(x)}_\mathrm{ss}$ is the ensemble average of $\lambda(x)$ over steady-state distribution $P_{\mathrm{ss}}(x)$. Therefore, Eq.~\ref{eq:first_order_stable} gives the first-order convergence condition, i.e, $|1-\eta \Braket{\lambda(x)}_\mathrm{ss}|<1$ or $\eta\Braket{\lambda(x)}_\mathrm{ss}<2$, equivalently. Expand $\lambda(x)$ around $x_\mathrm{ss}$, we get $\Braket{\lambda(x)}_\mathrm{ss}\approx \lambda(x_\mathrm{ss})+\frac{1}{2}\lambda''(x_\mathrm{ss})\left( \Braket{x^2}_\mathrm{ss} - x^2_\mathrm{ss}\right)$, which depends on the fluctuation $V_x$.

Moreover, we also require the variance of $y$ to be stable for SGD noise \cite{wu_how_2018}, the temporal average of $y^2$ can be approximated by:
\begin{equation} \label{eq:second_order_stable}
    \Braket{y^2_{t+1}}_t \approx  \left[ 1 + \eta^2 \Braket{\lambda^2(x)}_\mathrm{ss} - 2\eta \Braket{\lambda(x)}_\mathrm{ss}  + 2\eta^2 \sigma \Braket{\lambda'^2(x)} _\mathrm{ss} \right] \Braket{y^2_{t}}_t + 2\eta^2\sigma \Braket{\lambda^2(x)}_\mathrm{ss},
\end{equation}
which gives the second-order convergence condition, i.e., $ \eta^2 \Braket{\lambda^2(x)}_\mathrm{ss} - 2\eta \Braket{\lambda(x)}_\mathrm{ss} + 2\eta^2 \sigma \Braket{\lambda'^2(x)} _\mathrm{ss}<0$. $\Braket{\lambda^2(x)}_\mathrm{ss}$ and $\Braket{\lambda'^2(x)} _\mathrm{ss}$ also depends on the fluctuation $V_x$. From these two convergence stability conditions, we see that in the areas between Phase \uppercase\expandafter{\romannumeral2}\&\uppercase\expandafter{\romannumeral3} ($x_\mathrm{ss}\approx x_F$), large SGD noise level results in larger fluctuations, corresponding to higher probabilities on sharper locations, thus increases the fluctuation-dependent terms $\Braket{\lambda(x)}_\mathrm{ss}$, $\Braket{\lambda^2(x)}_\mathrm{ss}$ and $\Braket{\lambda'^2(x)} _\mathrm{ss}$. When the noise is large enough to reach the bounds of stability conditions, the system would finally diverge. Another noticeable point is that hyperparameters $\eta$ and $\sigma$ play different roles in the convergence conditions. To be clear, the fluctuation-dependent terms are typically determined by the product of hyperparameters $\eta\sigma$ (also see Sec.~\ref{sec:transition_boundary} for analytical results), while there is extra learning rate $\eta$ occurs in the convergence conditions. Therefore, increasing learning rate makes the system more likely to diverge, which explains the asymmetrical boundary in above simulations.  

\subsection{Numerical results of the critical exponents} \label{sec:critical_exponents}
\begin{figure}[htbp] 
    \includegraphics[width=0.98\textwidth]{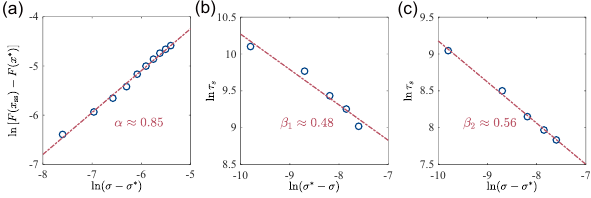}
    \caption{Fitting of the critical exponents $\alpha$, $\beta_1$ and $\beta_2$ while fixing learning rate $\eta=0.04$. The critical point $\sigma^{*}=0.0105$, and the landscape coefficients are the same as in Fig.2 of the main text. }
    \label{fig:critical_exponents}
\end{figure}

In the main text, we have observed a phase transition from the global minimum to flatter solutions (Phase  \uppercase\expandafter{\romannumeral1} to Phase  \uppercase\expandafter{\romannumeral2}) when increasing SGD noise. Here we'll show the numerical results of critical exponents of this phase transition. Since learning rate $\eta$ and noise strength $\sigma$ are nearly identical in this transition, without loss of generality, we'll show the case of fixing $\eta$ and changing $\sigma$. When flatness of steady-state solution $F(x_\mathrm{ss})$ serves as order parameter, there is a second-order transition near the critical point $\sigma^*$, where the flatness $F(x_\mathrm{ss})$ and noise strength $\sigma$ satisfies the power-law relation for $\sigma>\sigma^*$, i.e., $F(x_\mathrm{ss})-F(x^*)\sim(\sigma-\sigma^*)^{\alpha}$ where we defined a critical exponent $\alpha\approx0.85$, the fitting procedure is shown in Fig.~\ref{fig:critical_exponents}a. 

When search time $\tau_s$ serves as order parameter, the search time becomes infinite from both sides of the critical point, which is also known as the phenomenon of critical slowing down. This slowing down also follows a power-law from both form the left and right limits to critical point, thus we can define two critical exponents: $\tau_s \sim (\sigma - \sigma^{*})^{-\beta_{1}}$ for $\sigma>\sigma^{*}$ and $\tau_s \sim (\sigma^{*} - \sigma)^{-\beta_{2}}$ for $\sigma<\sigma^{*}$ with $\beta_{1}\approx 0.48$ and $\beta_{2}\approx 0.56$ as critical exponents, as shown in Fig.~\ref{fig:critical_exponents}b\&c. The numerical results here are consistent with the theoretical prediction by adiabatic elimination in Sec.~\ref{sec:critical_exponent}.

\section{Supplementary materials of analytical results}
\subsection{Continuous-time approximation of SGD updating rules} \label{sec:continuous_time}
To derive the continuous-time Langevin equations (Eq.~4 in the main text) from discrete SGD updating rules, we first rewrite the SGD updating rules in Eq.~1 as:
\begin{equation} \label{eq:SGD_updating_rules}
    \begin{split}
        \bm{\theta}_{t+1} &= \bm{\theta}_{t}-\eta \nabla_{\bm{\theta}} L\left(\bm{\theta}_{t}\right) + \eta \left[ \nabla_{\bm{\theta}} L^{\mu}\left(\bm{\theta}_{t}\right) - \nabla_{\bm{\theta}} L\left(\bm{\theta}_{t}\right) \right] \\
         & = \bm{\theta}_{t}-\eta \nabla_{\bm{\theta}} L\left(\bm{\theta}_{t}\right) + \sqrt{\eta}\bm{\xi}, 
    \end{split}
\end{equation}
where $\bm{\xi}\equiv \sqrt{\eta} \left[ \nabla_{\bm{\theta}} L^{\mu}\left(\bm{\theta}_{t}\right) - \nabla_{\bm{\theta}} L\left(\bm{\theta}_{t}\right) \right] $ is the rescaled SGD noise with covariance matrix $\eta\mathbf{C}$. Eq.~\ref{eq:SGD_updating_rules} can be viewed as Euler discretization of the following stochastic differential equation if we set the time interval $\dd{\tau} \equiv \eta$ when learning rate $\eta$ is small enough \cite{li_stochastic_nodate}:
\begin{equation} \label{eq:SDE}
    \dd{\bm{\theta}} =  - \nabla_{\bm{\theta}} L\left(\bm{\theta}\right) \dd{\tau} + \sqrt{\eta\mathbf{C}} \dd{W(\tau)},
\end{equation}
where $\dd{W(\tau)}$ is a standard N-dimensional wiener process, from this definition, the continuous-time unit $\tau$ is related with discrete iteration $t$ by learning rate $\eta$, i.e., $\tau=\eta t$ and that's the reason we rescaled search time by multiplying $\eta$. Mathematically, Eq.~\ref{eq:SDE} is equivalent to the following Langevin equations:
\begin{equation} \label{eq:Langevin}
    \dot{\bm{\theta}} =  - \nabla_{\bm{\theta}} L\left(\bm{\theta}\right)  + \bm{\xi},
\end{equation}
which is the general form of Eq.~4 in the main text. For simplicity, we redefine the noise covariance matrix of rescaled SGD noise $\bm{\xi}$ as diffusion matrix $\mathbf{D}$, i.e., $\mathbf{D}\equiv\eta\mathbf{C}/2$. Substituting the quadratic relation of $\mathbf{C}$ and $\mathbf{H}$, we obtain the expression of diffusion matrix $\mathbf{D}=\eta\sigma\mathbf{H}^2 = \Delta_S \mathbf{H}^2$ where we have used the definition of effective SGD noise level, $\Delta_S \equiv \eta\sigma$. Substituting Eq.~\ref{eq:Hessian_matrix} into the expression, we get the explicit formula for diffusion matrix:
\begin{equation} \label{eq:Diffusion_matrix}
    \mathbf{D}(x,y)=\Delta_S\begin{bmatrix}
        (\varepsilon\lambda_0 + \frac{1}{2}\lambda''(x)y^2)^2 + {\lambda'(x)}^2 y^2 & \lambda'(x)y \left( \varepsilon\lambda_0 + \lambda(x) + \frac{1}{2}\lambda''(x)y^2 \right) \\
        \lambda'(x)y \left( \varepsilon\lambda_0 + \lambda(x) + \frac{1}{2}\lambda''(x)y^2 \right) & \lambda(x)^2 + {\lambda'(x)}^2 y^2  \\
        \end{bmatrix} .
\end{equation}

\subsection{Analytical solutions of the Fokker-Planck equation by adiabatic elimination}

In this section, we derive the steady-state distribution $P_{\mathrm{ss}}(x,y)$ of Fokker-Planck equation (Eq.~11 in the main text) via adiabatic approximation. In general, nonlinear coupling of variables makes it challenging to solve the equation directly, however, due to the natural degeneracy of this problem, the timescales of these two variables can be separated under certain conditions thus enables us to solve it by adiabatic elimination of the fast variable. Specifically, the respective timescales are defined as the inverse of first-order partial derivatives of the drift terms of the Langevin equations in Eq.~4:
\begin{equation}{\label{eq:timescale}} 
    \tau_x = \left[ \varepsilon \lambda_0 + \frac{1}{2}\lambda''(x) y^2 \right]^{-1}   ; \quad
    \tau_y = \lambda(x)^{-1}.
\end{equation}

Besides $\varepsilon\ll 1$, the condition of timescale separation, i.e., $\tau_x \gg \tau_y$, requires relative small noise level $\Delta_S$, which will be discussed self-consistently in Sec.~\ref{sec:adiabatic_elimination}. For now, let's just assume these constraints are satisfied and continue the derivation. Under this assumption, the dynamics in $y$ direction relax much faster than in $x$ direction, which means given arbitrary $x$, $y$ has relaxed to stationary distribution while $x$ keeps nearly constant. Therefore, we can separate the $P_{\mathrm{ss}}(x,y)=P_{\mathrm{ss}}(x)P_{\mathrm{ss}}(y|x)$. If the cross-diffusion terms can be neglected (we also assume it holds true for now), $P_{\mathrm{ss}}(y|x)$ satisfies the following equations:  
\begin{equation} \label{eq:FP_y}
     \pdv{y} \left[  \lambda(x)  y  +  \pdv{y}  D_{22}(x,y)   \right] P_{\mathrm{ss}}(y|x) = 0,
\end{equation}
and $P_{\mathrm{ss}}(x)$ after adiabatic elimination follows one-dimensional Fokker-Planck equation under lowest order approximation \cite{kaneko_adiabatic_1981}: 
\begin{equation} \label{eq:FP_x}
    \dv{x} \left[  \varepsilon \lambda_0 x + \frac{1}{2} \lambda'(x)\Braket{y^2}_y  +  \dv{x}  \Braket{D_{11}(x,y)}_y   \right] P_{\mathrm{ss}}(x) = 0,
\end{equation}
where terms on $y$ has been integrated using conditional steady-state distribution $P_{\mathrm{ss}}(y|x)$, we use $\Braket{\cdot}_y\equiv\int{\cdot P_{\mathrm{ss}}(y|x)\dd{y}}$ to denote the integration. 

\subsubsection{$\Delta$ and $\kappa$ are constant}
For simplicity, we first consider the limiting case when $x\rightarrow 0$, which requires that SGD noise level is small enough. Under this limit, we see that the diffusion matrix (Eq.~\ref{eq:Diffusion_matrix}) is approximated by:
\begin{equation} \label{eq:Diffusion_matrix_approximated}
    \mathbf{D}(x=0,y)=\Delta_S\begin{bmatrix}
        (\varepsilon\lambda_0 + \lambda_1 y^2)^2  & 0 \\
        0 & \lambda_0^2 \\
        \end{bmatrix} ,
\end{equation}
where the cross-diffusion terms vanish, i.e., $D_{12} = D_{21}=0$ and $D_{22}$ becomes constant. We thus define the noise strength as the effective diffusion coefficient for $y$, i.e., 
\begin{equation}
    \Delta \equiv D_{22} = \Delta_S\lambda_0^2.   
\end{equation}
 Note that here the noise strength is a constant proportional to effective noise level $\Delta_S$. Substitute it into Eq.~\ref{eq:FP_y}, the solution $P_{\mathrm{ss}}(y|x)$ follows a Gaussian distribution:
\begin{equation} \label{eq:steady_state_x}
    P_{\mathrm{ss}}(y|x) = \sqrt{\frac{\lambda(x)}{ 2\pi \Delta}}\exp[-\frac{\lambda(x)}{2 \Delta} y^2] ,
\end{equation}
next, we define the anisotropy $\kappa$ as the ratio of effective diffusion coefficients between fast direction $y$ and slow direction $x$:
\begin{equation}
    \kappa \equiv \frac{\Delta}{\Braket{D_{11}}_y} \approx \frac{1}{\varepsilon^2 + 2\varepsilon\Delta_S\lambda_1 + 3\Delta_S^2\lambda_1^2},
\end{equation}
note here $\kappa$ is also a constant in the limit of $x\rightarrow 0$ and $\kappa$ is always larger than 1 for small effective noise level $\Delta_S$. However, increasing effective noise level would decrease the anisotropy. Then, substitute $\Braket{y^2}_y = \Delta /\lambda(x) $ and ${\Braket{D_{11}}_y}=\Delta / \kappa $ into Eq.~\ref{eq:FP_x}, we obtain the steady-state distribution $P_{\mathrm{ss}}(x)$:
\begin{equation} \label{eq:P_ss} 
    P_{\mathrm{ss}}(x) = \frac{1}{Z_{\mathrm{ss}}} \exp[-\frac{\kappa\varepsilon\lambda_0}{2\Delta} x^2] \lambda(x)^{-\frac{\kappa}{2}} ,
\end{equation}
where $Z_{\mathrm{ss}}\equiv \int P_{\mathrm{ss}}(x) \dd{x}$ is the normalization factor. Finally, we get the expression of steady-state distribution $P_{\mathrm{ss}}(x,y)$:
\begin{equation} 
    \begin{split}
        P_{\mathrm{ss}}(x,y) &= P_{\mathrm{ss}}(x)P_{\mathrm{ss}}(y|x) \\ &= \frac{1}{Z_{\mathrm{ss}}} \exp \left[-\frac{\kappa\varepsilon\lambda_0 x^2 + \lambda(x) y^{2}}{2 \Delta}\right] \lambda(x)^{-\frac{\kappa-1}{2}} ,       
    \end{split}
\end{equation}
where we have absorbed all the constants into normalization factor $Z_{\mathrm{ss}}$. As stated in the main text, the effective loss function can be defined from the steady-state distribution $P_{\mathrm{ss}}(x,y)$ where $\Delta$ plays the role as the effective temperature:
\begin{equation} \label{eq:effective_loss}
    \begin{split}
        L_\mathrm{eff}(x,y) &\equiv -\Delta \ln P_{\mathrm{ss}}(x,y) \\
        & = \frac{1}{2} \left[ \kappa\varepsilon\lambda_0 x^2 + \lambda(x)y^2 \right] + \Delta  (1-\kappa) \ln{F(x)} , \\      
    \end{split}
\end{equation}
where we have used the definition of flatness $F(x)\equiv 1/\sqrt{\lambda(x)}$ to replace $\lambda(x)$ and dropped all the constants. For completely degenerate case $\varepsilon = 0$, Eq.~\ref{eq:effective_loss} is exactly Eq.~7 of the main text. Compared with the expression of original loss function, we see that the additional loss introduced by SGD is given by:
\begin{equation} \label{eq:loss_SGD}
    \begin{split}
    L_\mathrm{SGD}(x) \equiv & L_\mathrm{eff}(x,y) - L(x,y) \\ = &  (\kappa - 1) \left[\frac{1}{2} \varepsilon\lambda_0 x^2 - \Delta \ln{F(x)} \right] \\
     = & \Delta (1 - \kappa) \ln{F(x)} +(\kappa-1)L(x,y=0),
    \end{split}
\end{equation}
recovering Eq.~11 of the main text.

\subsubsection{$\Delta$ and $\kappa$ are landscape-dependent}
For a more general case where the elements of diffusion matrix are landscape-dependent thus $\Delta$ and $\kappa$ can not be approximated as constants, the steady-state distribution $P_\mathrm{ss}(y|x)$ of Eq.~\ref{eq:FP_y} is given by the following distribution substituting $D_{22}= \lambda(x)^2 + {\lambda'(x)}^2 y^2$:
\begin{equation} \label{eq:P_y_general}
    P_\mathrm{ss}(y|x) \propto \left[ 1 + \frac{\lambda'(x)^2}{\lambda(x)^2} y^2 \right]^{-\frac{\lambda(x)}{2\Delta_S \lambda'(x)^2}-1},
\end{equation}
if the effective noise $\Delta_S$ satisfies $\Delta_S \lambda'(x)^2 \ll \lambda(x)$, we can define $\rho(x)\equiv \Delta_S \lambda'(x)^2/ \lambda(x)$, the distribution of Eq.~\ref{eq:P_y_general} can be approximated as the following Gaussian distribution:
\begin{equation}  
    \begin{split}
        P_\mathrm{ss}(y|x)  & \propto \lim_{\rho(x)\rightarrow 0}  \left[ 1 + \rho(x) \frac{y^2}{\Delta_S\lambda(x)} \right]^{-\frac{1}{2 \rho(x)}-1}  \\ &= \frac{1}{\sqrt{2\pi\Delta_S\lambda(x)}}\exp[-\frac{y^2}{2\Delta_S\lambda(x)} ], 
    \end{split}
\end{equation}
comparing with Eq.~\ref{eq:steady_state_x}, we see that under this assumption, the effective noise strength $\Delta(x)$ here is only dependent on $x$:
\begin{equation}
    \Delta(x)\equiv D_{22} \approx \Delta_S \lambda(x)^2. 
\end{equation}
Similar to previous section, we can define an effective anisotropy $\kappa(x)$ in a general form:
\begin{equation}
    \kappa(x) \equiv \frac{\Delta(x)}{\Braket{D_{11}(x,y)}_y} = \frac{\lambda(x)^2}{\varepsilon^2 \lambda_0^2 + \Delta_S[\varepsilon\lambda_0\lambda''(x)+ \lambda'(x)^2]\lambda(x)+\frac{3}{4}[\Delta_S\lambda''(x)\lambda(x)]^2},
\end{equation}
which is also the function of $x$. Substituting $\Braket{y^2}_y =  \Delta(x) / \lambda(x) $ and $\Braket{D_{11}(x,y)}_y = \Delta(x) / \kappa(x) $ into Eq.~\ref{eq:FP_x}, we obtain the Fokker-Planck equation after elimination:
\begin{equation} 
    \dv{x} \left[  \varepsilon \lambda_0 x +  \frac{\lambda'(x)\Delta(x)}{2\lambda(x)}  +  \dv{x}  \left( \frac{\Delta(x)}{\kappa(x)} \right)  \right] P_{\mathrm{ss}}(x) = 0,
\end{equation}
therefore, the general solution of $P_\mathrm{ss}(x)$ can be computed from the following integration:
\begin{equation} 
    P_\mathrm{ss}(x) = \frac{1}{Z_\mathrm{ss}}  \exp \left\{- \int{ \left[ \frac{\kappa(x)\varepsilon\lambda_0 x}{\Delta(x)} + \frac{\kappa(x)\lambda'(x)}{2\lambda(x)} \right] \dd{x}} \right\},
\end{equation}
which depends on the specific from of $\lambda(x)$. Finally, we have derived a general expression of steady-state solution $P_\mathrm{ss}(x,y)$:
\begin{equation} 
    \begin{split}
        P_{\mathrm{ss}}(x,y) &= P_{\mathrm{ss}}(x)P_{\mathrm{ss}}(y|x) \\ &= \frac{1}{Z_{\mathrm{ss}}} \sqrt{\frac{\lambda(x)}{\Delta(x)}} \exp \left\{ - \frac{\lambda(x)y^2}{2\Delta(x)} - \int{ \left[ \frac{\kappa(x)\varepsilon\lambda_0 x}{\Delta(x)} + \frac{\kappa(x)\lambda'(x)}{2\lambda(x)} \right] \dd{x}} \right\} ,       
    \end{split}
\end{equation}
similarly, we then obtain the general form of effective loss:
\begin{equation} \label{eq:effective_loss_general}
    \begin{split}
        L_\mathrm{eff}(x,y) &\equiv -\Delta(x) \ln P_{\mathrm{ss}}(x,y) \\ & = \frac{\lambda(x)y^2}{2} + \Delta(x)\int{ \left[ \frac{\kappa(x)\varepsilon\lambda_0 x}{\Delta(x)} + \frac{\kappa(x)\lambda'(x)}{2\lambda(x)} \right] \dd{x}}  -\frac{\Delta(x)}{2} \ln{\frac{\lambda(x)}{\Delta(x)}}, \\      
    \end{split}
\end{equation}
and the general form of SGD loss:
\begin{equation} 
    \begin{split}
    L_\mathrm{SGD}(x) \equiv & L_\mathrm{eff}(x,y) - L(x,y) \\ = & \Delta(x) \int{ \left[ \frac{\kappa(x)\varepsilon\lambda_0 x}{\Delta(x)} + \frac{\kappa(x)\lambda'(x)}{2\lambda(x)} \right] \dd{x}} -\frac{\Delta(x)}{2} \ln{\frac{\lambda(x)}{\Delta(x)} - \frac{1}{2}\varepsilon \lambda_0 x^2} ,
    \end{split}
\end{equation}
recovering Eq.~10 of the main text.

\subsubsection{Derivation of transition boundary} \label{sec:transition_boundary}
As we have shown in Fig.~2, for double-well form $\lambda(x)$, there is a phase transition from Phase \uppercase\expandafter{\romannumeral1} to Phase \uppercase\expandafter{\romannumeral2} when we increase the effective noise level $\Delta_S$ to a critical point. Analytically, the phase transition is the result of competition between original difference term and flatness-dependent term introduced by SGD. For small noise level, the original difference dominates thus the final solution is the global minimum; for large noise level, the flatness difference dominates thus the final solution becomes the flattest solution. Therefore, the phase-transition occurs when the global minimum $(x^*=0, y^*=0)$ starts to become saddle point, i.e., $\pdv[2]{L_\mathrm{eff}(x,y)}{x} \big|_{x=0,y=0}=0$. To obtain the critical condition, we first expand Eq.~\ref{eq:effective_loss_general} around $(0,0)$ to the second order:
\begin{equation}
    L_\mathrm{eff}(x,y) \approx \frac{\kappa(0)\lambda_0}{2} (\varepsilon + \Delta_S \lambda_1) x^2 + \frac{\lambda_0}{2}y^2,
\end{equation}
thus the transition occurs when:
\begin{equation}
    \pdv[2]{L_\mathrm{eff}(x,y)}{x} \bigg|_{x=0,y=0}= \frac{\kappa(0)\lambda_0}{2} (\varepsilon + \Delta_S \lambda_1) =0,
\end{equation}
which gives the transition boundary $\Delta_S = \Delta^{*}_S\equiv-\varepsilon/\lambda_1$ in the main text. 

As for the transition boundary $\Delta_m$ from convergence regime to divergence regime (Phase \uppercase\expandafter{\romannumeral2} to Phase \uppercase\expandafter{\romannumeral3}), we need to approximate $\Braket{\lambda(x)}_\mathrm{ss}$ to compute the convergence condition. For simplicity, we use steady-state distribution $P_\mathrm{ss}(x)$ of the constant case in Eq.~\ref{eq:P_ss}. We assume the noise level here is much larger than the transition boundary from Phase \uppercase\expandafter{\romannumeral1} to Phase \uppercase\expandafter{\romannumeral2}, i.e.,  $\Delta_S\gg\Delta_S^*$, which equals $|\Delta_S\lambda_1|\gg\varepsilon$, therefore:
\begin{equation}
    P_\mathrm{ss}(x)\approx \frac{1}{Z_\mathrm{ss}}\lambda(x)^{-\frac{\kappa}{2}},
\end{equation}
where anisotropy $\kappa\approx (3\Delta_S^2\lambda_1^2)^{-1}$. Next, $\Braket{\lambda(x)}_\mathrm{ss}$ can be approximated by:
\begin{equation} \label{eq:lambda_ss}
    \begin{split}
        \Braket{\lambda(x)}_\mathrm{ss} &= \int \lambda(x)P_\mathrm{ss}(x)\dd{x} = \frac{\int\lambda(x)^{-\frac{\kappa}{2}+1}\dd{x}}{\int\lambda(x)^{-\frac{\kappa}{2}}\dd{x}}\\
        & \approx \frac{\lambda(x_F)^{-\frac{\kappa}{2}+1}\int \exp[\frac{(\kappa+2)\lambda_1}{\lambda(x_F)}(x-x_F)^2]\dd{x}}{\lambda(x_F)^{-\frac{\kappa}{2}}\int \exp[\frac{\kappa\lambda_1}{\lambda(x_F)}(x-x_F)^2]\dd{x}} \\
        & = \lambda(x_F)\sqrt{\frac{\kappa+2}{\kappa}} \\
        & \approx \lambda(x_F)(1 + \frac{1}{\kappa}).
    \end{split},
\end{equation}
where we have used the approximation $\lambda(x)^{-\kappa/2}\approx \lambda(x_F)^{-\kappa/2}\exp[\frac{\kappa\lambda_1}{\lambda(x_F)}(x-x_F)^2]$ since $\kappa\gg1$. Substitute Eq.~\ref{eq:lambda_ss} into the first-order convergence condition $\eta\Braket{\lambda(x)}_\mathrm{ss}<2$, we have:
\begin{equation}
    \eta\lambda(x_F)(1+3\Delta_S^2\lambda_1^2)<2,
\end{equation}
therefore, the maximum $\Delta_m$ is given by:
\begin{equation}
   \Delta_m = \sqrt{\frac{1}{3\lambda_1^2}\left(\frac{2}{\eta_\mathrm{min}\lambda(x_F)}-1 \right)},
\end{equation}
where $\eta_\mathrm{min}$ is minimum learning rate, for $\eta_\mathrm{min}\ll 1$ and substitute the definition of flatness $F(x)$, we obtain the approximated boundary in the main text:
\begin{equation}
    \Delta_m \approx \sqrt{\frac{2}{3\eta_\mathrm{min}\lambda_1^2 }} F(x_F).
\end{equation}
Note that the analytical results of $\Delta_m$ are the upper bound of accurate $\Delta_m$ since we have computed under small noise approximation. The second-order convergence condition in Sec.~\ref{sec:fluctuation} gives the same bound for small noise. However, accurate $\Delta_m$ requires higher order expansion beyond small noise approximation, in that case the second-order convergence condition would give a stronger constraint than the first-order condition.

\subsubsection{Derivation of critical exponents} \label{sec:critical_exponent}
In this section, we derive the critical exponents $\alpha$, $\beta_1$ and $\beta_2$ defined in Sec.~\ref{sec:critical_exponents}. To compute the steady-state solution, we expand Eq.~\ref{eq:effective_loss_general} to the fourth order of variable $x$:
\begin{equation}
    L_\mathrm{eff}(x,y) \approx \frac{A}{2}(\Delta_S^{*} - \Delta_S) x^2 + \frac{B}{4} x^4 +  \frac{\lambda_0}{2}y^2,
\end{equation}
where we have used the definition of $\Delta_S^*$ in last section.  $A\equiv-\kappa(0)\lambda_0\lambda_1$ and $B\equiv \kappa(0) \Delta_S(\lambda_1^2 + 2\lambda_0\lambda_2)$ are both positive constants near the critical points. Therefore, the steady-state solution $x_\mathrm{ss}$ is given by the minimum of $L_\mathrm{eff}(x,y)$, i.e., $\pdv{L_\mathrm{eff}(x,y)}{x} \big|_{x_\mathrm{ss},y=0}=0$. When $\Delta_S < \Delta_S^*$, $x_\mathrm{ss}=x^*=0$; when $\Delta_S > \Delta_S^*$, $x_\mathrm{ss}$ is given by:
\begin{equation} \label{eq:x_ss}
    x_\mathrm{ss} = \pm \left[ \frac{A}{B} (\Delta_S - \Delta_S^*) \right] ^{\frac{1}{2}},
\end{equation}
thus the flatness of steady-state solution $F(x_\mathrm{ss})$ can be computed as:
\begin{equation} \label{eq:alpha}
    \begin{split}
        F(x_\mathrm{ss}) & \equiv \lambda(x_\mathrm{ss})^{-\frac{1}{2}} \\
        & \approx \lambda_0^{-\frac{1}{2}} (1 - \frac{\lambda_1}{2\lambda_0}x_\mathrm{ss}^2) \\
        &= F(x^{*}) - \frac{A\lambda_1}{2B \sqrt{\lambda_0^3}} (\Delta_S - \Delta_S^*),
    \end{split}
\end{equation}
where we have used the definition $F(x^*)=\lambda_0^{-\frac{1}{2}}$. From Eq.~\ref{eq:alpha}, we get the following power-law relation: $F(x_\mathrm{ss})-F(x^{*}) \sim (\Delta_S - \Delta_S^*)^{\alpha}$, the theoretical critical exponent $\alpha=1$.

When search time $\tau_s$ serves as order parameter, due to the timescale separation, we only need to consider the slow dynamics of $x$ direction, thus computation of the search time can be regarded as the one-dimensional first passage time problem governed by adiabatic Fokker-Planck equation in Eq.~\ref{eq:FP_x}. In particular, the average first passage time $\tau_s(x_0 \rightarrow x_\mathrm{ss})$ from starting point $x_0$ to steady-state solution $x_\mathrm{ss}$ is given by \cite{gardiner2009stochastic}:
\begin{equation} \label{eq:FPT}
    \tau_s(x_0 \rightarrow x_\mathrm{ss}) = \int_{x_\mathrm{ss}}^{x_0} \frac{\dd x}{P_\mathrm{ss}(x)} \int_{x}^{\infty} \frac{\kappa(x') P_\mathrm{ss}(x') \dd x' }{\Delta(x')},
\end{equation}
where we used the effective diffusion coefficient $\Braket{D_{11}(x,y)}_y=\Delta(x) / \kappa(x)$. Near the critical point, the integral is dominated by the contribution of vicinity of $x_\mathrm{ss}$, thus can be approximated by second order expansion. For $\Delta_S<\Delta_S^{*}$, $x_\mathrm{ss}=0$, and steady-state $P_\mathrm{ss}(x)\propto \exp [-A(\Delta_S^{*} - \Delta_S)x^2/2\Delta(0)]$, inserted into Eq.~\ref{eq:FPT}, we obtain:

\begin{equation} 
    \begin{split}    
        \tau_s(x_0 \rightarrow x_\mathrm{ss}) &= \frac{\kappa(0)}{\Delta(0)} \int_{0}^{x_0} \exp[\frac{A(\Delta_S^{*} - \Delta_S)x^2}{2\Delta(0)}] \dd x \int_{x}^{\infty} \exp[-\frac{A(\Delta_S^{*} - \Delta_S)x'^2}{2\Delta(0)}] \dd x'\\
        &\approx \frac{\kappa(0)}{\Delta(0)} \int_{0}^{x_0} \exp[\frac{A(\Delta_S^{*} - \Delta_S)x^2}{2\Delta(0)}] \dd x \left( \int_{0}^{\infty} \exp[-\frac{A(\Delta_S^{*} - \Delta_S)x'^2}{2\Delta(0)}] \dd x' \right)\\
        & \approx  \kappa(0)x_0 \left[ \frac{\pi}{2\Delta(0)A(\Delta_S^{*} - \Delta_S)} \right]^{\frac{1}{2}}, 
    \end{split}
\end{equation}
where we use the approximation $\int_0^{x_0}\exp[A(\Delta_S^{*} - \Delta_S)x^2/2\Delta(0)]\dd x \approx x_0$ near the critical point. Comparing with the definition of critical exponents, i.e.:
\begin{equation}
    \tau_s  \sim (\Delta_S^{*} - \Delta_S)^{-\beta_1}; \quad \tau_s  \sim (\Delta_S - \Delta_S^{*})^{-\beta_2},
\end{equation}
we finally derive the theoretical critical exponent $\beta_1=0.5$ for $\Delta_S<\Delta_S^{*}$. For the case when $\Delta_S>\Delta_S^{*}$, substitute the expression of steady-state $x_\mathrm{ss}$ in Eq.~\ref{eq:x_ss} and $P_\mathrm{ss}(x)\propto \exp [A(\Delta_S - \Delta_S^{*})(x-x_\mathrm{ss})^2/\Delta(0)]$ into Eq.~\ref{eq:FPT}, similarly, we obtain the theoretical critical exponent $\beta_2=0.5$. 

\subsubsection{Adiabatic elimination condition} \label{sec:adiabatic_elimination}
In this section we discuss the condition when adiabatic elimination (timescale separation) holds. The timescales defined in Eq.~\ref{eq:timescale} should satisfy $\tau_y \ll \tau_x$, we assume the condition holds and replace $y^2$ in $\tau_y$ with $\Braket{y^2}_y = T\lambda(x)$ to obtain a self-consistent condition:
\begin{equation} \label{eq:condition_general}
     \varepsilon\lambda_0 + \frac{\Delta_S}{2}\lambda''(x)\lambda(x) \ll \lambda(x),
\end{equation}
in the limit of $x\rightarrow0$, this condition can be simplified as:
\begin{equation} \label{eq:condition_zero}
     \varepsilon + \Delta_S\lambda_1 \ll 1,
\end{equation}
from Eq.~\ref{eq:condition_general} and Eq.~\ref{eq:condition_zero}, we see that the condition is associated with the effective noise level $\Delta_S$: when effective noise level is small, the steady-state probability is centered around zero, thus the adiabatic elimination assumption is largely met; when effective noise level becomes large, the steady-state probability becomes wider and since $\lambda''(x)$ increases with $x$, this condition becomes harder to be satisfied. 

In the main text we mentioned that a natural requirement in this question is that the average loss at steady-state should also be much less than 1, i.e., $\Braket{L}\approx\frac{1}{2}\Delta_S\lambda_0^2\ll 1$, which also ensures small $\Delta_S$. This condition is the same as the above adiabatic elimination condition when $\lambda_0$ and $\lambda_1$ are both on the order of $\mathcal{O}(1)$.

\begin{figure}[htbp] 
    \includegraphics[width=0.9\textwidth]{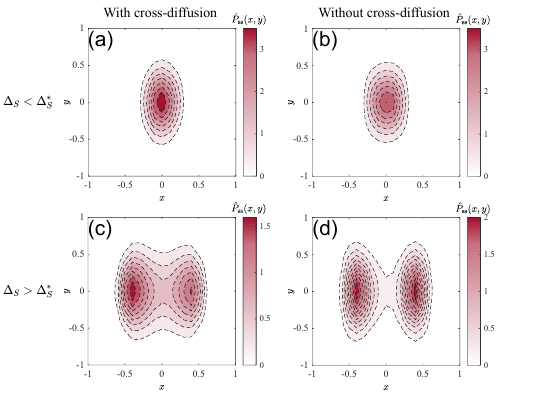}
    \caption{Estimated steady-state probability density function $\hat{P}_\mathrm{ss}(x,y)$ by simulation of SGD trajectories around steady-state. (a) Simulated $\hat{P}_\mathrm{ss}(x,y)$ with cross-diffusion terms fixing $\eta=0.01$ and $\sigma=0.9$. (b) Simulated $\hat{P}_\mathrm{ss}(x,y)$ without cross-diffusion terms fixing $\eta=0.01$ and $\sigma=0.9$. (c) Simulated $\hat{P}_\mathrm{ss}(x,y)$ with cross-diffusion terms fixing $\eta=0.01$ and $\sigma=1.2$. (d) Simulated $\hat{P}_\mathrm{ss}(x,y)$ without cross-diffusion terms fixing $\eta=0.01$ and $\sigma=1.2$. The landscape coefficients are the same as in Fig.~3 of the main text: $\varepsilon = 0.02$, $\lambda_0 = 8$, $\lambda_1 = -2$, $\lambda_2 = 1$ and transition boundary $\Delta_S^{*}=\varepsilon/\lambda_1=0.01$. }
    \label{fig:cross_diffusion}
\end{figure}

Another assumption we used in the above derivation is that the cross-diffusion terms can be neglected for small noise. Here, we verify this assumption by direct simulation of estimated steady-state probability density function $\hat{P}_\mathrm{ss}(x,y)$. As shown in Fig.~\ref{fig:cross_diffusion}, with or without cross-diffusion terms, the maximum locations and shapes of steady-state distribution are nearly the same, both for noise level below (Fig.~\ref{fig:cross_diffusion}a\&b) or beyond (Fig.~\ref{fig:cross_diffusion}c\&d) critical noise level $\Delta_S^{*}$. The simulation shows that the cross-diffusion terms has little effect on the final results thus can be neglected for adiabatic elimination approximation.

\subsection{Numerical solutions of the Fokker-Planck equation}
The stationary two-variable Fokker-Planck equation is solved numerically by PDE toolbox in MATLAB 2020b with finite element method. The solution region is a circular area of radius 2 with the center at $(0,0)$. We specify a Dirichlet condition $P_\mathrm{ss}(x,y)=0$ at the boundary of the solution region. We neglect the cross-diffusion terms in the diffusion matrix for numerical stability.

\clearpage
\bibliography{SGD}